\documentstyle[12pt]{article}

\newcommand {\m}{\mu}
\newcommand {\n}{\nu}
\newcommand {\pl}{\partial}
\newcommand {\p} {\phi}
\newcommand {\vp}{\varphi}

\newcommand {\al}{\alpha}

\newcommand {\ga}{\gamma}
\newcommand {\Ga}{\Gamma}

\newcommand {\la}{\lambda}

\newcommand {\si}{\sigma}

\newcommand {\e} {\mbox{\rm e}}
\newcommand {\na}{\nabla}
\newcommand {\del}  {\delta}

\newcommand {\mn}{{\mu\nu}}
\newcommand {\ls}   {{\lambda\sigma}}

\newcommand {\half}{ {\frac{1}{2}} }
\newcommand {\fourth} {\frac{1}{4} }
\newcommand {\sqg} {\sqrt{g}}
\newcommand {\fg}  {\sqrt[4]{g}}
\newcommand {\invfg}  {\frac{1}{\sqrt[4]{g}}}

\newcommand {\Dcal}{{\cal D}}
\newcommand {\Mcal}{{\cal M}}
\newcommand {\Ncal}{{\cal N}}
\newcommand {\Dvec}{{\vec D}}

\newcommand {\Ftil}{{\tilde F}}
\newcommand {\ftil}{{\tilde f}}

\newcommand {\intx} {{\int d^2x}}

\newcommand {\ra} {\rightarrow}
\newcommand {\pr}   {{\quad .}}
\newcommand {\com}  {{\quad ,}}
\newcommand {\q}    {\quad}

\newcommand {\nl}    {\newline}
\newcommand {\nn}    {\nonumber}
\newcommand {\vs}[1]  { \vspace*{#1 cm} }

\newcounter{eq}
\newcounter{sc}

\newcommand {\MPL}  {Mod.Phys.Lett.}
\newcommand {\NP}   {Nucl.Phys.}
\newcommand {\PL}   {Phys.Lett.}
\newcommand {\PR}   {Phys.Rev.}
\newcommand {\PRL}   {Phys.Rev.Lett.}

\newcommand {\CQG}  {Class.Quantum.Grav.}

\def\overleftrightarrow#1{\vbox{\ialign{##\crcr
 $\leftrightarrow$\crcr\noalign{\kern-1pt\nointerlineskip}
 $\hfil\displaystyle{#1}\hfil$\crcr}}}

\def\Dslash{{}\hbox{\hskip2pt\vtop
 {\baselineskip23pt\hbox{}\vskip-24pt\hbox{/}}
 \hskip-11.5pt $D$}}

\newlength{\minitwocolumn}
\begin{document}


\begin{flushright}
US-97-06\\
Febrary,1998(revised)\\
hep-th/9707025
\end{flushright}
\vspace{24pt}

\pagestyle{empty}
\baselineskip15pt

\begin{center}
{\large\bf Weyl Anomaly of 2D Dilaton-Scalar Gravity and \\
Hermiticity of System Operator \vskip 1mm
}

\vspace{10mm}

Shoichi ICHINOSE
          \footnote{
                    E-mail address:\ ichinose@u-shizuoka-ken.ac.jp
                  }
\\
\vspace{5mm}
          Department of Physics, University of Shizuoka,\\
          Yada 52-1, Shizuoka 422, Japan. \\
\end{center}

\vspace{15mm}
\begin{abstract}
Weyl(Conformal) anomaly in the dilaton-scalar system in 2 dimensional
gravity is examined. 
We take the heat-kernel regularization
for the ultraviolet divergences. 
Generally the Weyl anomaly is determined by
the 2nd order differetial (elliptic) operator of the system
and the definition of the measure. 
We have the freedom of the operator choice caused by
the arbitrariness of total divergences 
(surface terms) in the action.
We focus on such points
and the hermiticity of the system operator.
\vspace{15mm}

PACS NO:\ 11.10.-z,\ 04.60.K,\ 11.30.-j\nl
\end{abstract}

\newpage
\pagestyle{plain}
\pagenumbering{arabic}


\rm
\section{Introduction}

Motivated by 2 dim or 4 dim black hole physics\cite{CGHS}, 
Bousso and
Hawking\cite{BH} recently have calculated the Weyl anomaly
for the dilaton coupled scalar theory in 2 dim gravity.
\begin{eqnarray}
S_m = \half\int d^2 x \sqg \e^{-2\p}g^\mn \na_\m f\cdot\na_\n f\ ,
\label{1.1}
\end{eqnarray}
where $g_\mn,f$ and $\p$ are the metric, a scalar matter field and 
the dilaton respectively.
It is invariant under the Weyl transformation.
\begin{eqnarray}
\p'=\p\com\q f'=f\com\q g_\mn'=e^{2\al(x)}g_\mn\pr
\label{1.1b}
\end{eqnarray}
They have obtained, using the generalized zeta function,
the Weyl anomaly due to the scalar matter quantum effect
( $g_\mn$ and $\p$ are background fields ), as
\begin{eqnarray}
T= \sqg\left( q_1R+q_2(\na \p)^2+q_3\na^2\p\right),\nn\\
q_1=-\frac{1}{24\pi}\ ,\ \ q_2=-\frac{1}{4\pi}\ ,\ \ q_3=-\frac{1}{12\pi}\ .
\label{1.2}
\end{eqnarray}
They pointed out, in the induced effective action derived from the above
result, the term
$ -2\p R$ naturally appears from the $q_3$ term of eq.(\ref{1.2}). 
The term
was originally itroduced by hand in \cite{RST}
or from considerations of (Weyl) anomalies in \cite{BC}. 
The result shows
an important role of the
dilaton field. 
Stimulated by this work, some authors have examined
the problem in different ways.
The divergence structure of a wide class of theories, which includes
the above model, was already examined by \cite{ENO} in the dimensional
regularization.
Based on the work, Ref.\cite{NO} has calculated
the Weyl anomaly of (\ref{1.1}). Their result 
differs from eq.(\ref{1.2}) in $q_3=1/4\pi$. They have also obtained it for a wide
class of dilaton coupled scalar theories taking into account
not only the scalar matter but also the dilaton quantum effect.
Ref.\cite{MR,NO6143} has also examined the Weyl anomaly taking into
account all fields (graviton, dilaton and scalar matter) quantum effects.

It is true that the dilaton field is becoming more and more important
in accord with the rapid progress of the string theory. The string theory
predicts very naturally the existence of the dilaton in its low energy
(field theory) limit. 
We do not have much experience about the correct treatment of the
dilaton field and its characteristic interaction $\exp (-2\p)$. In this sense,
the close analysis of the present problem looks important.

Some problems in the Weyl anomaly itself also motivate us to do the present
analysis. Compared with the chiral anomaly ( anomaly related to "$\ga_5$" )
the general argument of Weyl anomaly is not so easy
\cite{BPB,DS93,AKMM,KMM}. 
It originates partly from the difficulty of 
clear characterization of Weyl-anomaly terms by a simple symmetry argument.
The Weyl anomaly looks to be more influenced by the local dynamics and 
generally requires
more complicated calculation than the chiral one. The present 2 dim model
is a nice simple model, by which we can test some general "belief" about
the Weyl anomaly. To say it more concretely, we focus on the following
points:\ 1) General structure of the anomaly terms;\ 2) Regularization
dependence of the Weyl anomaly;\ 3) The origin of the discrepancy
in the value of $q_3$;\ 4) How the arbitrariness of the total divergence
affects the Weyl anomaly. The quantization of the scalar field $f$ is
formally defined by the path-integral.
\begin{eqnarray}
\exp\,\Ga[g,\p]=\int d\m(f;g,\p)\,\exp\,S_m[f;g,\p]\com
\label{1.3}
\end{eqnarray}
where $d\m(f;g,\p)$ is the integration measure and requires the precise
definition. which will become a key point in later analysis.

There are
several ways of obtaining the Weyl anomaly:\ 
1) the generalized zeta function (\cite{BH});\ 
2) the counterterm formula expressed in the dimensional regularization(\cite{NO});\ 
3) the heat-kernel regularization(\cite{II96}).
We calculate the Weyl anomaly in the approach 3)
and take the Fujikawa's standpoint\cite{KF79,KF85} about the anomaly. 
Compared with the counter-term calculation with 
the dimensional regularization, the advantage of the present approach 
is that, once the differential operator of the system 
(we call it {\it system operator}) is fixed, 
we have no ambiguity, in all stages of calculation, 
arising from the total divergences\cite{foot2}. 
This is because
the heat-kernel approach deals with the operator directly, not with
the (effective) action. Ref.\cite{NO} hints that the discrepancy could
come from total divergence ambiguity in the counterterm
calculation. 
We will clarify this point in this paper.
Ref.\cite{NO} has also pointed out that the way of fixing $q_3$ which
ref.\cite{BH} has taken, is not unique because of the presence
of ambiguity     
in the procedure of translating
the Weyl anomaly result to the induced (nonlocal) action.

Before a detailed analysis, we fix the general form of the Weyl anomaly
by the symmetry arguments. Generally the anomaly is caused by 
a conflict between some symmetries. In the present case, 
they are the ( 2 dim )
general coordinate invariance (diffeomorphism) and the Weyl symmetry.
We take the familiar convention of keeping the general coordinate
invariance. Because of the quantization effect, 
the system cannot keep the Weyl symmetry. 
The general form is obtained as (\ref{1.2}) 
with general values of $q$'s by
the following requirements:
\ 1) 2 dim general coordinate invariance,;
\ 2) Weyl invariance (up to total divergences)\cite{BPB,DS93,AKMM,KMM};
\ 3) Invariance under
the constant shift of the dilaton field, $\p\ra \p+\mbox{const.}$.
The requirement 3) is necessary because the constant shift changes
the action (\ref{1.1}) only by an overall constant factor, which
can be absorbed into a constant rescaling of the scalar $f$. 
Without the condition 3) we would have
the arbitrariness of functionals $A(\p)$ and $B(\p)$ in
two terms, $A(\p)(\na\p)^2$ and $B(\p)\na^2\p$, 
as the anomaly general form\cite{foot3}. 
Among the three terms, $q_3$-term
is a "trivial term" in the terminology of the anomaly-cohomology
because the (infinitesimal) 
Weyl transformation of the local term $\sqg\p R$
shows
\begin{eqnarray}
(\sqg\p R)'=\sqg (\p R+2\al(x)\cdot \na^2\p)+\mbox{total-derivative}
+O(\al^2)\com\ \ |\al|\ll 1\pr \nn\\
\label{1.4}
\end{eqnarray}

\section{ Anomaly formula and hermiticity of system operator }

First we rescale the scalar field, $f$, in order to adjust the kinetic
term in the ordinary form: $f\exp (-\p)=F$. Then eq.(\ref{1.1}) is 
equivalent to, up to a total derivative, 
\begin{eqnarray}
& S_m^1= 
 \half\int d^2 x \ \fg F\Dvec_1 (\fg F)                    &  \nn\\
& = S_m-\half\int d^2 x \ \sqg \na_\m(F\na^\m F+a\na^\m\p\cdot F^2) \ ,& \nn\\
& \Dvec_1=\fg \left\{-\na^2+2(1-a)\na_\m\p\cdot\na^\m
-a\na^2\p+\na_\m\p\cdot\na^\m\p\right\}\invfg,  &
\label{2.1}
\end{eqnarray}
where a free parameter, $a$, appears in the {\it first} derivative term. 
It represents the
freedom of the total derivative terms. 
A purpose of this section is to check the effect of the 
freedom.
We can take the
BRS-invariant measure for the matter
scalar field by taking ${\tilde F}=\fg F$ instead of
$F$\cite{KF83}. This is the reason for the factors $\fg,\ 
\invfg$\ in $\Dvec_1$ above. 
The 2nd order differential
operator $\Dvec_1$ is hermitian  for $a=1$, when we
define the inner product and
the hermitian conjugate as
\begin{eqnarray}
&<{\tilde F_1}|{\tilde F_2}> =\intx {\tilde F_1}^\dag {\tilde F_2}
=\intx \sqg F_1^\dag F_2=\intx \sqg \e^{-2\p}f_1^\dag f_2\com &\nn\\
&<\Dvec_1{\tilde F_1}|{\tilde F_2}>
=<{\tilde F_1}|\Dvec_1^\dag {\tilde F_2}>. &
\label{2.1b}
\end{eqnarray}
This choice is most natural for the operator
$\Dvec_1$. For the case of $a\neq 1$, $\Dvec_1$ is not hermitian.
Note here that the factor of $\e^{-2\p}$ appears in the inner
product. This will be compared with the case of Sec.3.

The above definition of the inner product between the fields 
${\tilde F}$'s fixes the integration measure in the path integral
as follows\cite{KF79}. Let $\{u_n(x)\}$ be the complete
and orthogonal set of the eigen functions of the operator $\Dvec_1$.
\begin{eqnarray}
\Dvec_1u_n=\la_nu_n\com\q
<u_n|u_m>=\intx ~u_n^\dag u_m=\del_{nm}\pr
\label{2.A}
\end{eqnarray}
From the completeness, a general field ${\tilde F}=\fg F$ can be
expressed as
\begin{eqnarray}
{\tilde F}=\sum_n c_nu_n\com\q c_n\equiv <u_n|{\tilde F}>\com
\label{2.B}
\end{eqnarray}
where the orthogonality is used. Then the integration measure
in the path-integral can be precisely defined as
\begin{eqnarray}
&\exp\,\Ga_1[g,\p]=\int d\m_1(f;g,\p)\,\exp\,S_m^1[f;g,\p]\com\ \ 
S_m^1= \half\int d^2 x \Ftil \Dvec_1 \Ftil\com  &    \nn\\
&d\m_1(f;g,\p)=\prod_n dc_n \equiv\Dcal{\tilde F}&\pr
\label{2.C}
\end{eqnarray}
This measure $d\m_1$ depends on $a$ because $u_n$, by which the
expansion coefficients $c_n$ is obtained, depends on $a$ through
the eigen value equation (\ref{2.A}).

Following Fujikawa, let us evaluate the Jacobian $J$ in the
measure change due to Weyl transformation (\ref{1.1b}). We regard the Jacobian
as the Weyl anomaly.
\begin{eqnarray}
& \Ftil' \equiv\sum_n c_n'u_n 
 =\e^{\al(x)}\Ftil(x)=\sum_nc_n\e^{\al(x)}u_n\com &\nn\\
& c_n' =\sum_mc_m<u_n|\e^{\al(x)}u_m>
=c_n+\sum_mc_m<u_n|\al(x)u_m>+\mbox{O}(\al^2)\com &\nn\\
& J =\mbox{det}(\frac{\pl c_n'}{\pl c_m})
=\exp \mbox{Tr}\ln \left( \del_{mn}+<u_n|\al(x)u_m>+\mbox{O}(\al^2) \right) &\nn\\
& =\exp \left[\intx~\al(x)\sum_nu_n^\dag(x)u_n(x)+\mbox{O}(\al^2)\right].&
\label{2.D}
\end{eqnarray}
Therefore the Weyl anomaly is obtained as
\begin{eqnarray}
T_1=\left.\frac{\del J}{\del \al(x)}\right|_{\al=0}=\sum_n u_n^\dag(x)u_n(x)\pr
\label{2.E}
\end{eqnarray}
This expression is generally divergent, and must be regularized. We take
the heat-kernel regularization as
\begin{eqnarray}
&T_1 =\lim_{t\ra +0}\sum_n u_n^\dag(x)u_n(x)\e^{-\la_nt}
=\lim_{t\ra +0}\mbox{tr}~G_1(x,x;t)\com &\nn\\
&G_1(x,y;t) \equiv \sum_n u_n(x)\e^{-\la_nt}u_n^\dag(y)\pr &
\label{2.FF}
\end{eqnarray}
$G_1(x,y;t)$ satisfies the heat equation (\ref{2.G}) below
with $\Dvec=\Dvec_1$. For the case $a=1$, $\Dvec_1$ is
hermitian and we can write $T_1$ and $G_1(x,y;t)$ as
\begin{eqnarray}
&T_1 =\lim_{t\ra +0}\sum_n u_n^\dag(x)u_n(x)\e^{-\la_nt}
=\lim_{t\ra +0}\sum_n<n|x>\e^{-\Dvec_1t}<x|n> &\nn\\
&=\lim_{t\ra +0}\sum_n\mbox{tr}<x|n>\e^{-\Dvec_1t}<n|x>
=\lim_{t\ra +0}\mbox{tr}<x|\e^{-\Dvec_1t}|x>
=\lim_{t\ra +0}\mbox{tr}~G_1(x,x;t)\com &\nn\\
&G_1(x,y;t) \equiv <x|\e^{-\Dvec_1t}|y>\pr &
\label{2.F}
\end{eqnarray}

We note some general facts about 
the heat-kernel $G(x,y;t)$\cite{II96}. It
is rigorously defined by the heat equation in terms of 
an 2nd order differential operator $\Dvec$,
\begin{eqnarray}
\left(\frac{\pl}{\pl t}+\Dvec\right)~G(x,y;t)=0\pr
\label{2.G}
\end{eqnarray}
and can be generally solved perturbatively even for
the non-hermitian operator under the condition that the operator $\Dvec$
asymptotically goes to $-\del_\mn\pl_\m\pl_\n$
in the weak field limit\cite{II96}. In the present model,
the weak field means  $g_\mn=\del_\mn+h_\mn,\ 
|h_\mn|\ll 1,\ |\p|\ll 1$. 
We present a useful anomaly
formula valid for most 2 dim gravity-matter theories.
For a general theory, specified by general background fields
$\Ncal_\m$\ and $\Mcal$ ( which are general covariants) as below,
the Weyl anomaly $T$ is given by
\begin{eqnarray}
\Dvec=\fg (-\na^2-\Ncal_\m\na^\m-\Mcal)\invfg,\nn\\
T=\mbox{tr}~G(x,x;t)|_{t^0}
=\frac{1}{4\pi}\sqg~ \mbox{Tr}(-\frac{1}{6}R+\Mcal-\half\na_\m\Ncal^\m).
\label{2.2}
\end{eqnarray}
"Tr" in the above equation means "trace" over all field indices:
\ $i,j=1,2,\cdots,N;\ \mbox{Tr}\Mcal=\Mcal^{ii}, \mbox{etc}$. 
This formula is easily derived from two facts:\ 
1) the formula must be general invariant, 
2) for the weak field gravity case, it leads 
to eq.(27) of ref.\cite{II96} or eq.(40) of ref.\cite{DAMTP9687}.

In the present special case of $\Dvec_1$ (
$\Ncal_\m=-2(1-a)\na_\m\p,\ \Mcal=a\na^2\p-\na_\m\p\cdot\na^\m\p,\ 
N=1$),
we obtain the Weyl anomaly as
\begin{eqnarray}
T_1=\frac{1}{4\pi}\sqg\left(-\frac{1}{6}R-\na_\m\p\cdot\na^\m\p
+\na^2\p\right)\com\nn\\
q_1=-\frac{1}{24\pi}\com\q q_2=-\frac{1}{4\pi}\com\q
q_3=\frac{1}{4\pi}\pr
\label{2.3}
\end{eqnarray}
The final result does not depend on the free parameter $a$\, 
and agrees with that of Ref.\cite{NO}
.\cite{foot4}
The result of $a$-independence
say the arbitrariness of total derivative terms in the
{\it first} derivative does not affect the Weyl anomaly.
\section{ Second Choice of Measure  }
The dilaton-scalar theory (\ref{1.1}) can be modified to,
up to total divergences,
\begin{eqnarray}
S_m^2 = \half\intx \sqg \e^{-2\p}f(-\na^2+2\na_\m\p\cdot\na^\m)f \nn\\
=S_m+\half\intx\sqg\na^\n(\e^{-2\p}f\na_\n f)\pr
\label{3.1}
\end{eqnarray}
$S_m^2$ seems to give the same Weyl anomaly as $S_m^1$ if the situation
is the same as in Sect.2. In this section we take the following one 
\begin{eqnarray}
\Dvec_2\equiv \fg \e^{-2\p}(-\na^2+2\na_\m\p\cdot\na^\m)\frac{1}{\fg}\com
\label{3.2}
\end{eqnarray}
as the system operator. The operator $\Dvec_2$ 
is hermitian only for the inner product and the hermitian
conjugate defined by
\begin{eqnarray}
&\ftil \equiv\fg f\com &\nn\\
&<\ftil_1|\ftil_2> \equiv \intx \ftil_1^\dag(x)\ftil_2(x)
=\intx\sqg f_1^\dag(x) f_2(x)\com &\nn\\
&<\Dvec_2\ftil_1|\ftil_2> =<\ftil_1|\Dvec_2^\dag\ftil_2>\pr &
\label{3.3}
\end{eqnarray}
In contrast with the inner product (\ref{2.1b}) in Sect.2,  
the dilaton factor $\exp(-2\p)$ does not appear in the above
definition. The factor appears in the operator $\Dvec_2$.
As in the Sect.2, the path-integral measure must be taken as
\begin{eqnarray}
& \Dvec_2v_n=\si_nv_n\com\q
<v_n|v_m>=\intx~ v_n^\dag v_m=\del_{nm}\com  & \nn\\ 
& {\tilde f}=\sum_n b_nv_n\com\q b_n\equiv <v_n|{\tilde f}>\com & \nn\\
& \exp\,\Ga_2[g,\p]=\int d\m_2(f;g,\p)\,\exp\,S_m^2[f;g,\p]\com
S_m^2= \half\int d^2 x \ftil \Dvec_2 \ftil\com      & \nn\\
& d\m_2(f;g,\p)=\prod_n db_n\equiv\Dcal{\tilde f}\pr &
\label{3.3b}
\end{eqnarray}
The same procedure as given in Sect.2 leads to the Weyl anomaly, using the
measure defined by (\ref{3.3}), 
\begin{eqnarray}
&T_2=\lim_{t\ra +0}\sum_n v_n^\dag(x)v_n(x)\e^{-\si_nt}
=\lim_{t\ra +0}\sum_n<n|x>\e^{-\Dvec_2t}<x|n> &\nn\\
&=\lim_{t\ra +0}\sum_n\mbox{tr}<x|n>\e^{-\Dvec_2t}<n|x>
=\lim_{t\ra +0}\mbox{tr}<x|\e^{-\Dvec_2t}|x>
=\lim_{t\ra +0}\mbox{tr}~G_2(x,x;t)\com &\nn\\
&G_2(x,y;t)\equiv <x|\e^{-\Dvec_2t}|y>\pr&
\label{3.4}
\end{eqnarray}

We note the anomaly formula (\ref{2.2}) can not be applied for
$\Dvec_2$ because of the dilaton factor $\exp(-2\p)$ in front of
the {\it second} derivative term.
We evaluate $G_2(x,x;t)$ in the following two steps.
\flushleft{i) The flat space}  \nl
In the case of the flat space, $g_\mn=\del_\mn$, $\Dvec_2$ reduces to
\begin{eqnarray}
\Dvec_2= \e^{-2\p}(-\pl^2+2\pl_\m\p\cdot\pl_\m)\pr
\label{3.5}
\end{eqnarray}
Now we consider the weak field case of the dilaton:\ $|\p|\ll 1$. We fix
the anomaly in a first few orders with respect to $\p$. $\Dvec_2$ is 
expressed as
\begin{eqnarray}
&\Dvec_2 = -\pl^2-W_\mn\pl_\m\pl_\n-N_\m\pl_\m-M\com &\nn\\
&W_\mn =(\e^{-2\p}-1)\del_\mn=\left(-2\p+2\p^2+\mbox{O}(\p^3)\right)
\del_\mn\com &\nn\\
&N_\m =-2\e^{-2\p}\pl_\m\p=-2(1-2\p)\pl_\m\p+\mbox{O}(\p^3)\com &\nn\\
&M =0\com&
\label{3.6}
\end{eqnarray}
where $W_\mn,N_\m$ and $M$ are introduced for the general 
background fields 
and their explicit content, in the present case, are given above.
From the definition of the heat-kernel (\ref{2.G}), $G_2$ satisfy
\begin{eqnarray}
\left(\frac{\pl}{\pl t}-\pl^2\right)~G_2(x,y;t)=
(W_\mn\pl_\m\pl_\n+N_\m\pl_\m+M)G_2(x,y;t)\com
\label{3.6b}
\end{eqnarray}
which is used to solve the heat-kernel perturbatively with respect
to the fields in the right-hand side $(W,N,M)$. The 1-st order result of
$G_2(x,x;t)|_{t^0}$ was obtained in (27) of \cite{II96} as
\begin{eqnarray}
\frac{1}{4\pi}(-\frac{1}{12}\pl^2W_{\m\m}+\frac{1}{3}\pl_\m\pl_\n W_\mn
-\half\pl_\m N_\m+M)\pr
\label{3.7}
\end{eqnarray}
This formula gives
\begin{eqnarray}
\frac{1}{4\pi}(\frac{2}{3}\pl^2\p-\frac{4}{3}(\pl\p)^2)\pr
\label{3.8}
\end{eqnarray}
The 2nd order formula is similarly obtained as\cite{foot5}
\begin{eqnarray}
&\frac{1}{4\pi} \{
\frac{1}{48}\pl_\m W_{\la\la}\cdot\pl_\m W_{\si\si}
-\frac{1}{12}\pl_\m W_\mn\cdot\pl_\n W_{\la\la}
-\fourth\pl_\m W_\mn\cdot\pl_\la W_{\la\n} &\nn\\
&+\frac{1}{24}\pl_\m W_\ls\cdot\pl_\m W_\ls
+\frac{1}{12}\pl_\m W_{\n\la}\cdot\pl_\la W_\mn
-\fourth N_\m N_\m+\half\pl_\m W_\mn\cdot N_\n
                                            \} &\nn\\
			   &+(W\pl\pl W,W\pl N,WM \mbox{terms})\pr&
\label{3.9}
\end{eqnarray}
This formula gives, for the present model,
\begin{eqnarray}
\frac{1}{4\pi}\times \frac{1}{3}(\pl\p)^2\pr
\label{3.10}
\end{eqnarray}
Therefore we have
\begin{eqnarray}
q_2=
\frac{1}{4\pi}(-\frac{4}{3}+\frac{1}{3})=-\frac{1}{4\pi}\com\q
q_3=\frac{1}{4\pi}\times\frac{2}{3}\pr
\label{3.11}
\end{eqnarray}
\flushleft{ii) Vanishing Dilaton}  \nl 
In the case of vanishing dilaton, $\p=0$, $\Dvec_2$ reduces to
$-\fg\na^2\invfg$, and the Weyl anomaly is well known ( for example,
see (29) of \cite{II96}).
\begin{eqnarray}
q_1=
-\frac{1}{24\pi}\pr
\label{3.12}
\end{eqnarray}
We note, in (\ref{3.11}), that 
the value of $q_3$ differs from the result in Sect.2\cite{foot6}
whereas $q_2$ is the same.
The $q_3$ value agrees with (16) of Ref.\cite{KLV} with the choice
$\psi(\p)=0$ (trivial measure) and $\vp(\p)=\p$ in the notation
of the reference where the zeta-function regularization is taken.

\section{ Discussion and Conclusion }
Generally the way to analyse a quantum field theory using some background
(external) fields is quite popular in the theoretical physics. A systematic
approach is called the background field method\cite{BD67,GtH73} or
the effective action approach\cite{BOS92}. The anomaly formulae (\ref{2.2})
,(\ref{3.7}) and (\ref{3.9})
(or its 4 dim version in \cite{II96,DAMTP9687}) are based on the
formalism. It should be noted that the anomaly formula is different
from the counter-term formula in that it is definite including
total derivative terms. 

\q Taking the heat-kernel regularization,
we have analysed the Weyl anomaly
with special interst in the freedom of the 
choice of the system operator, 
which is caused by the arbitrariness of the total divergence. 
The operator $\Dvec_2$ of Sect.3
was taken in \cite{BH}, whereas $\Dvec_1$ of Sect.2 was essetially taken
in \cite{NO}. Although they are related by the total divergence, the corresponding
Weyl anomalies become different because of the measure difference. 
(From this viewpoint, it is considered that the $a$-independence
of the Weyl anomaly in Sect.2 shows that the measure difference
due to the {\it first} derivative term do not influence the Weyl anomaly.)
The most natural measures are fixed for each case from the requirement
of the hermiticity of the system operators.

\q Let us compare the present situation with the case of the non-Abelian
anomaly\cite{WB69,KF79,KF85}, where typically two types of anomalies
exist, i.e., the consistent anomaly and the covariant anomaly.
In Ref.\cite{DAMTP9687}, they are treated in the present approach.
For the system of the chiral (right) fermion coupled with the gauge
field in 4 dim space:\ $
{\cal L}_R = - {\bar\psi} \gamma^\mu ( \partial_\mu 
- i\frac{1+\gamma_5}{2} R_\mu) \psi
\equiv - {\bar\psi} \Dslash_{R+} \psi
$, 
the consistent anomaly is obtained by the choice 
$\Dvec_{cons}=-\Dslash_{R+}\Dslash_{R+}$(non-hermitian), 
and the covariant one by
$\Dvec_{cov}=\Dslash_{R+}^\dag\Dslash_{R+}$(hermitian). 
$\Dvec_{cons}$ 
is directly related with the effective
action as
$\sqrt{\det\Dvec_{cons}}=\int\Dcal\psi\Dcal{\bar\psi}\e^{\int{\cal L}_R}$. 
On the other hand, $\Dvec_{cov}$ does not have
such a direct relation:\ 
$\det\Dvec_{cov}=|\int\Dcal\psi\Dcal{\bar\psi}\e^{\int{\cal L}_R}|^2$. 
This corresponds to the fact that the
consistent anomaly satisfies the consistency condition derived
from the "formal"(without touching on the regularization problem)  
effective action analysis, whereas the covariant one
does not. In the Fujikawa's standpoint, however, the anomaly is not 
defined
by the "formal" effective action analysis. 
It originates from the 
{\it regularization}
of $\del^n(x-y)$ (in $n$ dim space) 
which appears in the Jacobian at the change
of the integration variable fields.  The anomaly
comes 
from the integration measure at the {\it regularized} level. 
In the heat-kernel  regularization, we take 
$\del^n(x-y)\equiv\lim_{t\ra +0}<x|\e^{-\Dvec t}|y>$
where the connection with the system is provided by
the operator $\Dvec$. This "flexibility" is one of the advantages
of the Fujikawa's approach.  
In this standpoint, the regularization dependence in
the Weyl anomaly found in the present paper is similar to
that in the non-Abelian anomaly except the following points.
\begin{enumerate}
\item $\Dvec_1$ and $\Dvec_2$ in the Weyl anomaly case 
are related by the total divergences,
whereas $\Dvec_{cons}$ and $\Dvec_{cov}$ in the non-Abelian anomaly 
case are not.
\item Corresponding to the above item, 
in the Weyl anomaly case the two choices of the operators
give the difference only
in a part of the whole anomaly, that is, the trivial term.
In the non-Abelian anomaly case, the two choices give
totally different anomalies.   
\end{enumerate}

\q The heat-kernel calculation in Sect.3 is equivalent to 
the generalized
zeta function ($\zeta_{\Dvec_2}(s)$) analysis taken in \cite{BH}. It is well known
$\zeta_{\Dvec_2}(s)$ is related with $G_2(x,x;t)$ of (\ref{3.4}) as
\begin{eqnarray}
Y(t)\equiv\intx~ G_2(x,x;t)=\sum_n\e^{-\si_nt}\com\nn\\
\zeta_{\Dvec_2}(s)=\frac{1}{\Ga(s)}\int_0^{\infty}Y(t)t^{s-1}dt
=\sum_n\si_n^{-s}\pr
\label{4.1}
\end{eqnarray}
Technically Ref.\cite{BH} evaluated $\zeta_{\Dvec_2}(s)|_{s=0}$,
whereas this paper has evaluated $G_2(x,x;t)|_{t=0}$.

\q We have treated the metric and the dilaton as background (classical) fields.
Their quantum effects can be straightforwardly taken into account by
following the background formalism. (In fact they are evaluated in 
\cite{NO,MR,NO6143}.)
At present, however, it looks that the correct quantum treatment
of the metric and the dilaton will be given
by the string theory. 

\q The Weyl anomaly is a key factor in the Hawking radiation phenomenon.
In the present case of 2 dimension, the energy momentum tensor $T_\mn$
is determined by the conformal anomaly and the conservation equation
with an appropriate boundary condition.
\begin{eqnarray}
&\sqg g^\mn T_\mn = \sqg\left( -\frac{1}{24\pi}R
-\frac{1}{4\pi}(\na \p)^2+q_3\na^2\p\right) ,&\nn\\
&\na^\m T_\mn=0&\pr
\label{4.2}
\end{eqnarray}
The different values of $q_3$ should not affect physical quantities.
If the above equations are right, they show 
the energy momentum tensor itself
depends on $q_3$. It seems to give some trouble about
the physical interpretation of the energy momentum tensor
and the radiation amplitude.   
We should notice here the following points:\ 
1)\ we do not take into account
the contribution of the quatum effect of the dilaton and the graviton;\ 
2)\ the $q_3$ term in (\ref{4.2}) can be freely changed by introducing a 
finite counterterm, $\p R$, 
in the gravity-dilaton lagrangian in advance as noted in (\ref{1.4}).
This problem looks to involve graviton-dilaton dynamics
and "renormalization" of its lagrangian. 
In the present analysis of perturbative quantization of matter fields,
we have no control over such a problem.
Although the above problem is beyond the
scope of the present work, it is highly interesting that 
the black hole physics is not closed within the quantum matter-field
theory on the background curved space, but open to the quantum gravity 
problem through the Weyl anomaly.

\vs 1
\begin{flushleft}
{\bf Acknowledgement}
\end{flushleft}
The author thanks Prof.S.D.Odintsov 
and Prof.S.Nojiri for informing their
stimulating results and discussions about this subject. 
He also thanks Prof. K.Fujikawa
for an encouraging comment and some corrections in the original 
manuscript.

\vs 1


\begin{thebibliography}{99}
\bibitem{foot3}
{We do not take the 'parity' symmetry $\p\ra -\p$, which appears
in the S-duality in the string theory. That demands $q_3=0$, which
is inconsistent with the explicit (perturbative) result.}
\bibitem{foot4}
{The result $q_3=1/4\pi$ also agrees with (16) of Ref.\cite{KLV}
with $\vp(\p)=\psi(\p)=\p$ in its notation.}
\bibitem{foot5}
{Eq.(\ref{3.9}) is obtained by evaluating eq.(25) of Ref.\cite{II96}
taking $n=2$ there.}
\bibitem{foot6}
{$q_3$ of (\ref{3.12}) differs from \cite{BH} too.
Note that their derivation of $q_3$
is ambiguous as mentioned in Sect.1.
Because the calculational circumstance (regularization, operator,
measure) is essentially the same (see Sect.4), we expect their value
of $q_3$ should coincide with (\ref{3.11}).}
\end{thebibliography}
\end{document}